\documentclass[conference]{IEEEtran}

\usepackage[OT1]{fontenc} 
\usepackage[numbers,sort&compress]{natbib}
\usepackage[cmex10]{amsmath}
\usepackage{amssymb}
\usepackage{bm}
\usepackage{graphicx}
\usepackage{color}
\usepackage{subfigure}
\usepackage{tabularx}
\usepackage{arydshln}
\usepackage{mathtools}
\usepackage{multirow}
\usepackage{algorithm,algorithmic}
\usepackage{url}
\usepackage{listings}
\usepackage{comment}
\usepackage{hyperref}
\usepackage[absolute,overlay]{textpos}

\newcommand{\argmin}{\mathop{\rm arg~min}\limits}
\def\I{\mathbf{I}}
\def\e{\mathbf{e}}
\def\E{\mathbf{E}}
\def\U{\mathbf{U}}
\def\u{\mathbf{u}}
\def\y{\mathbf{y}}
\def\Y{\mathbf{Y}}

\def\Yh{\hat{\mathbf{Y}}}

\def\H{\mathbf{H}}

\def\S{\mathbf{S}}

\def\v{\mathbf{v}}
\def\V{\mathbf{V}}

\def\X{\mathbf{X}}
\def\Xh{\hat{\mathbf{X}}}

\def\A{\mathbf{A}}
\def\M{\mathbf{M}}

\makeatletter
\def\ps@IEEEtitlepagestyle{%
  \def\@oddfoot{\mycopyrightnotice}%
  \def\@oddhead{\hbox{}\@IEEEheaderstyle\leftmark\hfil\thepage}\relax
  \def\@evenhead{\@IEEEheaderstyle\thepage\hfil\leftmark\hbox{}}\relax
  \def\@evenfoot{}%
}
\def\mycopyrightnotice{%
  \begin{minipage}{\textwidth}
  \centering \scriptsize
\textcopyright 2021 IEEE. Personal use of this material is permitted.
  Permission from IEEE must be obtained for all other uses, in any current or future
  media, including reprinting/republishing this material for advertising or promotional
  purposes, creating new collective works, for resale or redistribution to servers or
  lists, or reuse of any copyrighted component of this work in other works.
  DOI: \href{https://doi.org/10.1109/VTC2021-Fall52928.2021.9625337}{10.1109/VTC2021-Fall52928.2021.9625337}
  \end{minipage}
}
\makeatother

\begin{document}
\title{Optimal but Low-Complexity Optimization Method for Nonsquare Differential Massive MIMO}

\author{\IEEEauthorblockN{Yuma Katsuki\IEEEauthorrefmark{1} and Naoki Ishikawa\IEEEauthorrefmark{1}}
\IEEEauthorblockA{\IEEEauthorrefmark{1}Graduate School of Engineering Science, Yokohama National University, 240-8501 Kanagawa, Japan.}}

\markboth{}
{Shell \MakeLowercase{\textit{et al.}}: Bare Demo of IEEEtran.cls for Journals}
\maketitle
\TPshowboxesfalse
\begin{textblock*}{\textwidth}(45pt,10pt)
\footnotesize
\centering
This is the author's final version which has not been fully edited and content may change prior to final publication. Citation information: \href{https://doi.org/10.1109/VTC2021-Fall52928.2021.9625337}{10.1109/VTC2021-Fall52928.2021.9625337}
\end{textblock*}
\begin{abstract}
In this paper, we propose an optimal but low-complexity optimization method for nonsquare differential massive MIMO. While a discrete nonlinear optimization is required for the conventional nonsquare differential coding, we newly modify it to perform a low-complexity continuous linear optimization. This novel method exhibits immediate convergence as compared to the conventional method. Additionally, the proposed differential coding can be regarded as a differential counterpart of the coherent generalized spatial modulation. Our numerical comparisons demonstrate that the proposed method achieves the best coding gain for any number of transmit antennas, although the optimization cost is nearly negligible.
\end{abstract}


\IEEEpeerreviewmaketitle

\section{Introduction}
Automated vehicles have to rely on wireless communications in order to ensure safety.
In vehicle-to-vehicle and vehicle-to-infrastructure communications, we need to maintain a reliable connection even at night, in heavy rainfall and/or high mobility scenarios \cite{kong_millimeter-wave_2017}. If there is a high latency or error in the transmission of data, it could cause a serious traffic accident.
Thus, wireless technology with lower latency and higher reliability is indispensable for future networks.

Spatial modulation (SM) \cite{mesleh2008spatial} 
has been proposed as a multiple-input multiple-output (MIMO) scheme to simplify both transmitter and receiver while maintaining the same transmission rate as the conventional spatial multiplexing.
SM conveys information by selecting one of $M$ transmit antennas and improves the transmission rate on the order of $\mathrm{log}_2(M)$.
Later, the generalize SM (GSM) \cite{jeganathan_generalized_2008} is proposed to further improve the transmission rate.
Both SM and GSM require accurate channel state information (CSI) at the receiver. The effect of rapid changes in the radio propagation environment cannot be ignored when assuming communication with high-mobility scenarios, such as cars in automatic operation.

Square-matrix-based differential space-time coding (S-DSTC) \cite{hughes_differential_2000,tarokh_differential_2000} does not require CSI estimation and insertion of pilot symbols.
A bit sequence is mapped to an $M \times M$ unitary matrix, and this data matrix is transmitted using $M$ time slots with $M$ antennas.
It is effective for environments where CSI changes rapidly; however, it imposes a large burden for massive MIMO scenarios since it inevitably requires a large number of time slots \cite{ishikawa2018pm}.

Nonsquare-matrix-based differential space-time coding (N-DSTC) \cite{ishikawa_rectangular_2017,xiao2020differentiallyencoded,wu2018rdsm} has been studied since 2017 that is capable of reducing the number of transmit time slots from $M$ to $T~(T<M)$. 
This scheme is especially suitable for millimeter-wave communications in high-speed mobile environments \cite{ishikawa_differential-detection_2019}. 
It is demonstrated in \cite{ishikawa_differential_2018} that the nonsquare counterpart of the diagonal unitary coding (DUC) \cite{hochwald_differential_2000} achieves competitive performance, although it requires a high-complexity discrete optimization.
The time complexity of this DUC optimization becomes severe as the number of transmit antennas increases.

Against this background, we propose an optimal but low-complexity optimization method for the N-DSTC scheme. The conventional discrete optimization problem required for N-DSTC is newly transformed into a continuous optimization problem, which significantly reduces time complexity.
Additionally, the proposed differential scheme can be regarded as a differential counterpart of the GSM scheme.
Our numerical comparisons demonstrate that the proposed scheme is scalable to an increase in the number of transmit antennas.

\section{Conventional Differential MIMO\label{sec:sys} \cite{rajashekar_algebraic_2017,hochwald_differential_2000}}
In this paper, we assume a MIMO system with $M$ transmit antennas and $N$ receive antennas, where the number of transmit antennas is a power of two, i.e., $M=2^1,2^2, \cdots$.
The received symbol block is given by
\begin{align}
\Y(i) = \H(i)\S(i) + \V(i) \in \mathbb{C}^{N \times M}
\label{COH:eq:blockmodel}
\end{align}
where $i$ denotes a transmission index, $ \H(i) \in \mathbb{C}^{N \times M}$ denotes a channel matrix that obeys the i.i.d. Rayleigh fading $\mathcal{CN}(0,1)$, and $ \S(i) \in \mathbb{C}^{M \times M}$ denotes a space-time codeword. The codeword $\S(i)$ is transmitted by $M$ antennas over $M$ time slots. We assume that the additive noise $ \V(i) \in \mathbb{C}^{N \times M}$ follows the i.i.d complex Gaussian distribution, $\mathcal{CN}(0,\sigma_v^2)$, and the signal-to-noise ratio (SNR) is calculated as $1/\sigma_v^2$, i.e., $10 \cdot \mathrm{log}_{10}(1/\sigma_v^2)$ [dB].

The S-DSTC \cite{hughes_differential_2000,tarokh_differential_2000} scheme has been proposed as a MIMO scheme that does not require estimation of CSI. This scheme maps $B$ [bit] input bit sequence to a data matrix $\X(i)\in\mathbb{C}^{M \times M}$ and multiplies the data matrix $\X(i)$ by the previous block $\S(i-1)$, which is called differential encoding.
Finally, the differentially-encoded block $\S(i)$ is transmitted through $M$ antennas over $M$ time slots.

We use two representative S-DSTC schemes: the algebraic differential spatial modulation (ADSM) \cite{rajashekar_algebraic_2017} and the diagonal unitary coding (DUC) \cite{hochwald_differential_2000}.

The ADSM scheme \cite{rajashekar_algebraic_2017} maps $B$ [bit] information to the data matrix $\X(i)$ as follows.
First, the $B$-length input bit sequence is partitioned into two sequences: $B_1=\log_2(M)$ [bit] and $B_2=\log_2(L)$ [bit].
The first $B_1$ [bit] information is used for selecting a specific dispersion matrix (DM) out of $M$ DMs  $\A_1,\cdots, \A_M \in \mathbb{C}^{M \times M} $, which are given by 
\begin{align}
    \{\A_1, \cdots, \A_M\} = \{\I_M, \M, \M^2, \cdots, \M^{M-1}\}.
    \label{COH:eq:ADSM:batrix}
\end{align}
Here, the permutation matrix $\M$ is defined by
\begin{align}
	\M=\left[\begin{array}{ccccc}
	0 & 0 & \cdots & 0 & e^{j 2\pi/L} \\
	1 & 0 & \cdots & 0 & 0\\
	0 & 1 & \cdots & 0 & 0\\
	\vdots & \vdots & \ddots & \vdots & \vdots\\
	0 & 0 & \cdots & 1 &0
	\end{array}\right]\in \mathbb{C}^{M \times M}.
	\label{COH:eq:permutation:matrix}
\end{align}
The selected DM is represented as $\A_m(i)$.
Second, the $B_2$ [bit] information is mapped to an $L$-PSK symbol $s(i)\in\mathbb{C}$.
Finally, a data matrix $\X(i)=s(i)\A_m(i)$ is generated.

By contrast, the DUC scheme \cite{hochwald_differential_2000} maps $B$ [bit] information to the data matrix $\X(i)$ as follows.
The $B$-length input bit sequence is mapped to a decimal number $b$ and the corresponding data matrix is defined by
\begin{align}
\X(i) = \mathrm{diag}\left[\mathrm{exp}\left(j\frac{2\pi b}{2^B}u_1\right), \cdots, \mathrm{exp}\left(j\frac{2\pi b}{2^B}u_M\right)\right].
\label{COH:DUC:detamatrix}
\end{align}
Here, the $M$ number of diversity-maximizing factors $0 < u_1 \leq \cdots u_M \leq2^B/2 \in \mathbb{Z}$ are designed so as to maximize the diversity product of
\begin{align}
P_{max}=\min_{b\in\{1, \cdots, 2^B-1\}}\left|\prod_{m=1}^M \mathrm{sin}\left(\frac{\pi b u_m}{2^B}\right)\right|^\frac{1}{M}.
\label{COH:DUC:fanc}
\end{align}

The search space size of \eqref{COH:DUC:fanc} can be calculated as ${2^{(B-1)}+M-1 \choose M}$.
For example, we have ${3M-1 \choose M}$ candidates for $B=\mathrm{log}_{2}(M)+2$.
Then, if we use $M=(32,64,128,256\cdots)$ transmit antennas, we have to consider $(2\times10^{25},5\times10^{51},4\times10^{104},4\times10^{210}\cdots)$ candidates, respectively.
Hence, the optimization of DUC causes a combinatorial explosion.

After the data matrix $\X(i)$ is generated, the corresponding space-time codeword is differentially encoded by
\begin{align}
    \S(i) = \left\{ \begin{array}{ll}
    \I_M & (i=0) \\
    \S(i-1)\X(i) & (i>0)
  \end{array} \right.
\label{COH:differencial}
\end{align}
The noncoherent maximum likelihood (ML) detection for $i>0$ is carried out by:
\begin{align}
\Xh(i)=\argmin_{\X} \|\Y(i)-\Y(i-1)\X \|^2_F.
\label{COH:nonsq:detection}
\end{align}

\section{Conventional Nonsquare Coding \cite{ishikawa_differential_2018}\label{sec:conv}}

The space-time codeword $ \S(i) \in \mathbb{C}^{M \times M} $ of S-DSTC is transmitted in $M$ time slots using $M$ antennas. Increasing the number of transmit antennas results in increasing the number of required time slots, which is inefficient for massive MIMO scenarios. Therefore, nonsquare-matrix-based DSTC (N-DSTC) \cite{ishikawa_rectangular_2017,ishikawa_differential_2018} has been proposed to reduce the number of required time slots.
Specifically, it multiplies $\S(i)$ by the basis $\e_1\in \mathbb{C}^{M \times 1}$ to reduce the number of time slots from $M$ to $1$.

\subsection{Algebraic Construction Method for Nonsquare Basis\label{subsec:basis}}
The N-DSTC scheme relies on a basis $\e_1\in \mathbb{C}^{M \times 1}$.
The basis has to satisfy the following constraints for $1\leq k\neq k' \leq M$: 
\begin{align}
\text{Power constraint}:&~ \|\e_k\|^2_F=1
\label{COH:power:constration}\\
\text{Orthogonality}:&~ \e^H_k\e_k=1 \ \text{and} \  \e^H_k\e_{k'}=0
\label{COH:orthogonality}\\
\text{Reconstructability}:&~ \sum_{k=1}^{M} \e_k\e^H_k=\I_M
\label{COH:reconstructability}
\end{align}
In general, the component sub-matrices taken from a unitary matrix $\U_M \in \mathbb{C}^{M \times M}$ satisfy all the above constraints. Thus, we construct a set of bases as $[\E_1\ \E_2\  \cdots \ \E_{M/T}]=\U_M$.

We introduce a specific construction method for $\e_1$. With $M$ transmit antennas and $N_b$ non-zero components in each column, the corresponding basis is given by
\begin{align}
\U_M=\mathrm{bdiag}\underbrace{[\mathbf{W}, \cdots, \mathbf{W}]}_{M/N_b~\text{repetition}}\in \mathbb{C}^{M \times M}.
\label{COH:hybrid:basis}
\end{align}
In \eqref{COH:hybrid:basis}, the discrete Fourier transform matrix $\mathbf{W}$ is expressed by
\begin{align}
	\mathbf{W} = \frac{1}{\sqrt{N_b}}\left[\begin{array}{cccc}
	1 & 1 & \cdots & 1\\
	1 & \omega & \cdots & \omega^{N_b-1}\\
	1 & \omega^2 & \cdots & \omega^{2(N_b-1)}\\
	\vdots & \vdots & \ddots & \vdots\\
	1 & \omega^{N_b-1} & \cdots  &\omega^{(N_b-1)(N_b-1)}
	\end{array}\right],
	\label{COH:DFT:matrix}
\end{align}
where we have $\omega=\mathrm{exp}(-2\pi j/N_b)$.
When $N_b=1$ and $N_b=M$, both cases are specially called sparse basis and dense basis, respectively.
Since $M/N_b$ has to be an integer, we use $N_b=1,2^1,2^2,\cdots, 2^{\mathrm{log_2}(M)}$ in this paper.

\subsection{Nonsquare Differential Encoding and Decoding\label{subsec:nonsq_encording}}
The N-DSTC scheme transmits the reference symbols of $\{\e_1, \cdots, \e_{M}\}$ over $M$ time slots, which is the same as the conventional differential family. Thus, during the block index $1 \leq i \leq M$, the received symbol block is expressed as 
\begin{align}
\y(i)=\H(i)\e_i+\v(i) \in \mathbb{C}^{N \times 1}.
\label{COH:transmit:pilot}
\end{align}
The data matrix $\X(i)$ for $M<i \leq W$ is differentially encoded according to \eqref{COH:differencial}, and then the space-time codeword $\S(i)$ multiplied by $\e_1$ is transmitted.
Thus, the received symbol block is expressed as
\begin{align}
\y(i)=\H(i)\S(i)\e_1+\v(i).
\label{COH:transmit:data}
\end{align}
The reference insertion ratio is defined by $\eta=M/W$ and we set $W=20M$ so that $\eta$ becomes 5\%

The noncoherent ML detection for $i > M$ is carried out by
\begin{align}
\Xh(i)=\argmin_{\X} \|\y(i)-\Yh(i-1)\X\e_1 \|^2_F
\label{COH:nonsq:detection}
\end{align}
where $\Yh(i)$ is defined by 
\begin{align}
    \Yh(i) = \left\{ \begin{array}{ll}
    \sum_{k=1}^{M} \y(k)\e^H_k & (i=M) \\
    \y(i)\e^{(1-\alpha)}+\Yh(i-1)\Xh(i)\E^{(\alpha)} & (i>M)
  \end{array} \right.
\label{COH:Y:hat}
\end{align}
In \eqref{COH:Y:hat}, $\E^{(\alpha)}$ and $\e^{(1-\alpha)}$ are given by 
\begin{align}
    \left\{ \begin{array}{ll}
    \E^{(\alpha)}=\alpha(i) \e_1 \e_1^H + \sum_{k=2}^{M} \e_k \e_k^H \\
    \e^{(1-\alpha)}=(1-\alpha(i))\e_1^H .
  \end{array} \right.
\label{COH:constant:matrix}
\end{align}
In \eqref{COH:constant:matrix}, the adaptive forgetting factor $\alpha(i)$ is given by \cite{ishikawa_differential-detection_2019}
\begin{align}
    \alpha(i)=\min\left[\max\left[\frac{N \cdot \sigma_v^2}{\|\mathbf{d}(i) \|^2_F},0.01\right],0.99\right],
\label{COH:forgetting:factor}
\end{align}
where we have $\mathbf{d}(i)=\y(i)-\Yh(i-1)\Xh(i)\e_1$.

\section{Proposed Nonsquare Coding\label{sys:pnon}}
To maximize performance, we clarify a minimum requirement to pursue the performance upper bound.
Specifically, we generalize the non-zero elements value of the basis $\e_1$ to complex numbers.
In addition, we propose a communication method that can achieve the same performance as the conventional nonsquare differential scheme even if we ignore the two constraints that were considered necessary.

\subsection{High-performance Basis via Continuous Optimization\label{subsec:prop:basis}}
The conventional study \cite{ishikawa_differential_2018} claims that the basis $\e_1$ must satisfy \eqref{COH:power:constration}--\eqref{COH:reconstructability}.
In this paper, we reveal that the only condition that must be satisfied is only \eqref{COH:power:constration}. Therefore, we analyze the performance of $\e_1$ that satisfies the power constraint of \eqref{COH:power:constration}, and clarify novel conditions for achieving the performance limit. We first consider the case of $M=N_b$ and extend it to the case of arbitrary $N_b$.
When $\e_1\in \mathbb{C}^{M \times 1}$ satisfies the power constraint and the power is equally distributed among all the transmit antennas, $\e_1$ is given by
\begin{align}
\e_1=[e_1\ e_2\ \cdots\ e_M]^T / \sqrt{M} \in \mathbb{C}^{M \times 1}.
\label{COH:propose:basis}
\end{align}
Here, each element $e_m\ (1 \leq m \leq M)$ of $\e_1$ lies on a circle of radius one in the complex number plane and is represented by $e_m=e^{j\theta_m}(0 \leq \theta_m <2\pi)$. 

As a metric to evaluate the basis, we use the coding gain expressed by \cite{hanzo_near-capacity_2009}
\begin{align}
g(\e_1)=\min_{\X_1 \neq \X_2} \left|(\X_1\e_1-\X_2\e_1)^H(\X_1\e_1-\X_2\e_1) \right|^\frac{1}{N},
\label{COH:coding:gain}
\end{align}
which calculates the minimum Euclidean distance between $\X_1 \e_1$ and $\X_2 \e_1$.
To calculate \eqref{COH:coding:gain}, we have to take any two matrices from $\X_1, \cdots, \X_{2^B} \in \mathbb{C}^{M \times M}$, calculate the Euclidean distance, and find the minimum.
Hence, we have to consider ${2^B \choose 2}$ matrices to calculate \eqref{COH:coding:gain} to evaluate a basis $\e_1$. 
For example, in the $B=8$ [bit] case, ${2^8 \choose 2}=32640$ matrices are considered, which is not suitable for massive MIMO scenarios with a large number of transmit antennas.

When we use DUC, the coding gain does not depend on the value of the basis.
By contrast, when we use ADSM and calculate the coding gain, we obtain the following simplified expression.
\begin{align}
    g(\e_1)=\min(g_1(\e_1), g_2(\e_1)),
    \label{COH:min:coding:gain}
\end{align}
where we have
\begin{align}
     g_1(\e_1)&=\min_{s_1 \neq s_2} 
    \left|s_1-s_2\right|^\frac{2}{N} \mathrm{and}
    \label{COH:propose:G1}\\
      g_2(\e_1)&=\min_{\substack{n=1, \cdots, M/2}} \left[2-2 \mathrm{Re}\{s \cdot\e_1^H \M^n \e_1\} \right]^\frac{1}{N}.
    \label{COH:propose:G2}
\end{align}
Note that $s_1$ and $s_2$ are arbitrary $L$-PSK symbols used to represent the $B_2$-length information.
If we limit $N=1$, the maximum value of $g(\e_1)$ is 2 for both $B_2=1$ and $B_2=2$ cases.
The coding gain $g$ can be maximized when we have $| \e_1^H \M^n \e_1| =0$ for all $n$ with $1 \leq n \leq M/2$.
Based on this observation, we propose a novel objective function of
\begin{align}
f(\e_1)=\sum_{n=1}^{M/2}|\e_1^H \M^n \e_1|.
\label{COH:propose:fanc}
\end{align}
For example, in the $M=2$ case, we obtain the optimal solution $\e_1=[1~e^{j\frac{\pi}{4}}]^T/\sqrt{2}$, which contains a complex value unlike the conventional N-DSTC.

In \eqref{COH:propose:fanc}, we need to perform $M/2$ matrix calculations to evaluate one basis, which is much smaller than the original case \eqref{COH:coding:gain}.
Specifically, when we have $M=64$ and $B=8$, the number of matrix calculations required to evaluate one basis in \eqref{COH:coding:gain} can be reduced from 32640 to 32.

\subsection{Extension for the $N_b < M$ case}
In Section~\ref{subsec:prop:basis}, the number of nonzero components in $\e_1$ is limited to $M$.
Here, we extend the proposed basis to support an arbitrary number of nonzero components, which is denoted by $N_b \leq M$.
Specifically, the original basis $\e_1$ of \eqref{COH:propose:basis} is represented as $\e_1(M, M)$ since it is designed for $M$ transmit antennas and has $M$ nonzero components.
We newly define $\e_{1} (M,N_b) \in \mathbb{C}^{M \times 1}$ as the optimal basis for the $M$ and $N_b~(< M)$ case.
The extended basis $\e_{1} (M,N_b) \in \mathbb{C}^{M \times 1}$ for $M > N_b$ can be designed by a recursive manner as follows:
\begin{align}
\e_1(M, N_b)=\e_1 \left(\frac{M}{2}, N_b \right) \otimes \begin{bmatrix}1 \\ 0\end{bmatrix} \in \mathbb{C}^{M \times 1}.
\label{COH:expand:basis}
\end{align}
For example, if we consider the $M=4$ and $N_b=2$ case, the corresponding extended basis is calculated as
\begin{align}
\e_1(4, 2) &= \e_1(2, 2) \otimes \begin{bmatrix}1 \\ 0\end{bmatrix}
= \frac{1}{\sqrt{2}} \begin{bmatrix}1 \\ e^{j\frac{\pi}{4}}\end{bmatrix} \otimes \begin{bmatrix}1 \\ 0\end{bmatrix}
=\frac{1}{\sqrt{2}} \begin{bmatrix} 1 \\ 0 \\ e^{j\frac{\pi}{4}} \\ 0 \end{bmatrix}.
\end{align}
If we have $B=4$, the $4 \times 4$ ADSM codewords are projected by $\e_1(4,2)$ as follows:
\begin{align}
    \frac{1}{\sqrt{2}} \left\{\begin{bmatrix} 1 \\ 0 \\ e^{j\frac{\pi}{4}} \\ 0\end{bmatrix},\begin{bmatrix} 0 \\ 1 \\ 0 \\ e^{j\frac{\pi}{4}} \end{bmatrix},
    \begin{bmatrix} e^{j\frac{3\pi}{4}} \\ 0 \\ 1 \\ 0 \end{bmatrix},
    \begin{bmatrix} 0 \\ e^{j\frac{3\pi}{4}} \\ 0 \\ 1 \end{bmatrix},
    \begin{bmatrix} j \\ 0 \\ e^{j\frac{3\pi}{4}} \\ 0 \end{bmatrix},\cdots
    \right\}.
\end{align}
which are similar to the GSM codewords \cite{jeganathan_generalized_2008}.

\subsection{Generalized Transmission Procedure \label{subsec:propose_encording}}
The conventional N-DSTC scheme has to prepare $\e_2, \cdots, \e_M$ that satisfies the constraints \eqref{COH:power:constration}--\eqref{COH:reconstructability}, in addition to the generated basis $\e_1$.
Although there is a systematic method to generate $\e_2, \cdots, \e_M$ \cite{ishikawa_differential_2018}, the calculation cost may not be ignored when the number of transmit antennas is large.
In this paper, we propose a novel N-DSTC scheme that relies on $\e_1$ only.

Our proposed transmission procedure does not require $\e_2, \cdots, \e_M$.
Instead, we generate a random unitary matrix that is shared between the transmitter and the receiver.
Specifically, an arbitrary unitary matrix $\U_M=[\u_1, \cdots, \u_{M}]\in \mathbb{C}^{M \times M}$ is generated, where we have $\U_M=\I_M$ in the simplest case.
First, the unitary matrix $\U_M=[\u_1, \cdots, \u_{M}]$ is transmitted in $M$ time slots as reference symbols. The received symbol block for $1 \leq i \leq M$ is expressed as
\begin{align}
 \y(i)=\H(i)\u_i+\v(i).
 \label{COH:propose:pilot}
\end{align}
Then, we use the basis $\e_1\in \mathbb{C}^{M \times 1}$ for data transmission.
The data matrix $\X(i)$ for $M<i \leq W$ is differentially encoded according to \eqref{COH:differencial}, and the space-time codeword $\S(i)$ is mapped onto an $M \times 1$ rectangular matrix as follows:
 \begin{align}
 \y(i)=\H(i)\S(i)\e_1+\v(i).
 \label{COH:propose:sender}
 \end{align}

\subsection{Simplified ML Detection\label{subsec:propose_detection}}
We propose a noncoherent ML detector that corresponds to Section~\ref{subsec:propose_encording} as follows:
\begin{align}
\Xh(i)=\argmin_{\X} ||\y(i)-\Yh(i-1)\X\e_1||^2_F
\label{COH:propose:nonsq:detection}
\end{align}
where $\Yh(i)$ is defined by 
\begin{align}
    \Yh(i) = \left\{ \begin{array}{ll}
    \sum_{k=1}^{M} \y(k)\u^H_k & (i=M) \\
    \y(i)\e^{(1-\alpha)}+\Yh(i-1)\Xh(i)\E^{(\alpha)} & (i>M)
  \end{array}. \right.
\label{COH:propose:Y:hat}
\end{align}
In \eqref{COH:propose:Y:hat}, we have $\E^{(\alpha)}=\I_M-\e_1\e^{(1-\alpha)}$ and $\e^{(1-\alpha)}=(1-\alpha(i))\e_1^H$.
The definition of $\alpha(i)$ is the same as that used in \eqref{COH:forgetting:factor}.
The proposed detector \eqref{COH:propose:nonsq:detection} does not include $\e_2, \cdots, \e_M$, which can further simplify the N-DSTC scheme.

\section{Performance Comparisons\label{sec:comp}}
We evaluate the complexity of the proposed optimization method and its performance in terms of average mutual information (AMI) and bit error ratio (BER) by Monte Carlo simulations.

\begin{figure}[tb!]
	\centering
	\includegraphics*[clip,scale=0.64]{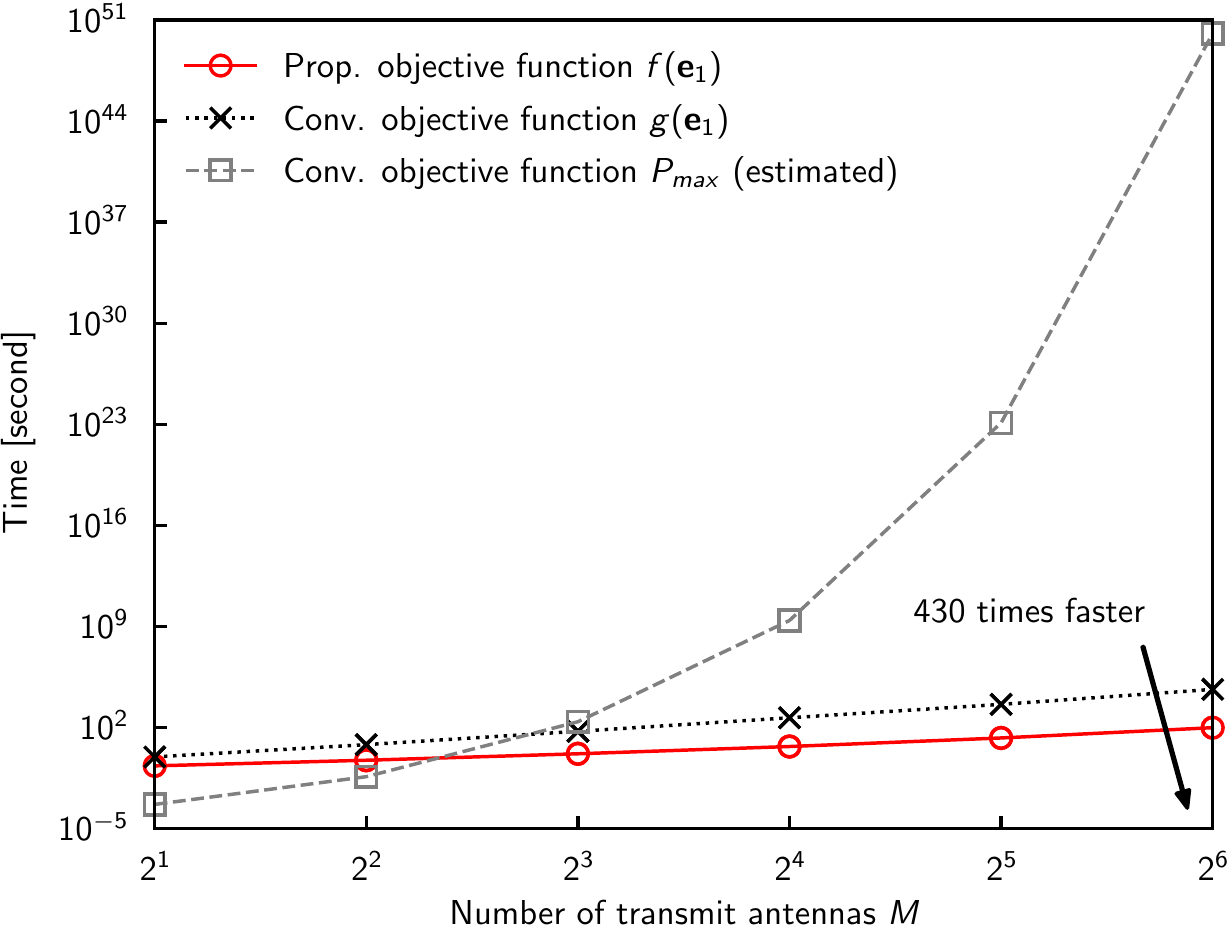}
	\caption{Effective time required to complete the conventional and proposed optimization methods, where the number of transmit antennas was varied from $M=2$ to $64$ and we set the length of input bit sequence $B=\mathrm{log}_2(M)+2$.
	\label{fig:compare:time}}
\end{figure}
Fig.~\ref{fig:compare:time} shows the effective time required to complete the conventional and proposed optimization methods, where we had $M=2,4,\cdots,64$ and $B=\mathrm{log}_2(M)+2$. 
The discrete nonlinear optimization is required for the DUC scheme, and its completion time is too large to be verified.
Instead, in the DUC case, we used completion time expected from a small number of iterations.\footnote{We generated a set of $u_1, u_2, \cdots ,u_M$ and measured its real evaluation time. The completion time can be expected by multiplying it by the search space size.}
For the optimization of $g(\e_1)$ defined in \eqref{COH:coding:gain} and $f(\e_1)$ defined in \eqref{COH:propose:fanc}, we used the dual annealing method \cite{xiang_generalized_1997} that was capable of obtaining a global optimal solution.
As shown in Fig.~\ref{fig:compare:time}, in the $M=2^{6}=64$ case, the DUC required about $3 \times 10^{43}$ years to complete the optimization, which was a reference.
The conventional objective function, $g(\e_1)$, required $43000$ seconds to complete the dual annealing optimization, while the proposed counterpart, $f(\e_1)$ required $100$ seconds, which was $430$ times faster.
This gap monotonically increases as the number of transmit antennas increases since complexity order of $g(\e_1)$ and $f(\e_1)$ are $O(M^3)$ and $O(M^2)$, respectively.

\begin{figure}[t]
	\centering
	\includegraphics*[clip,scale=0.64]{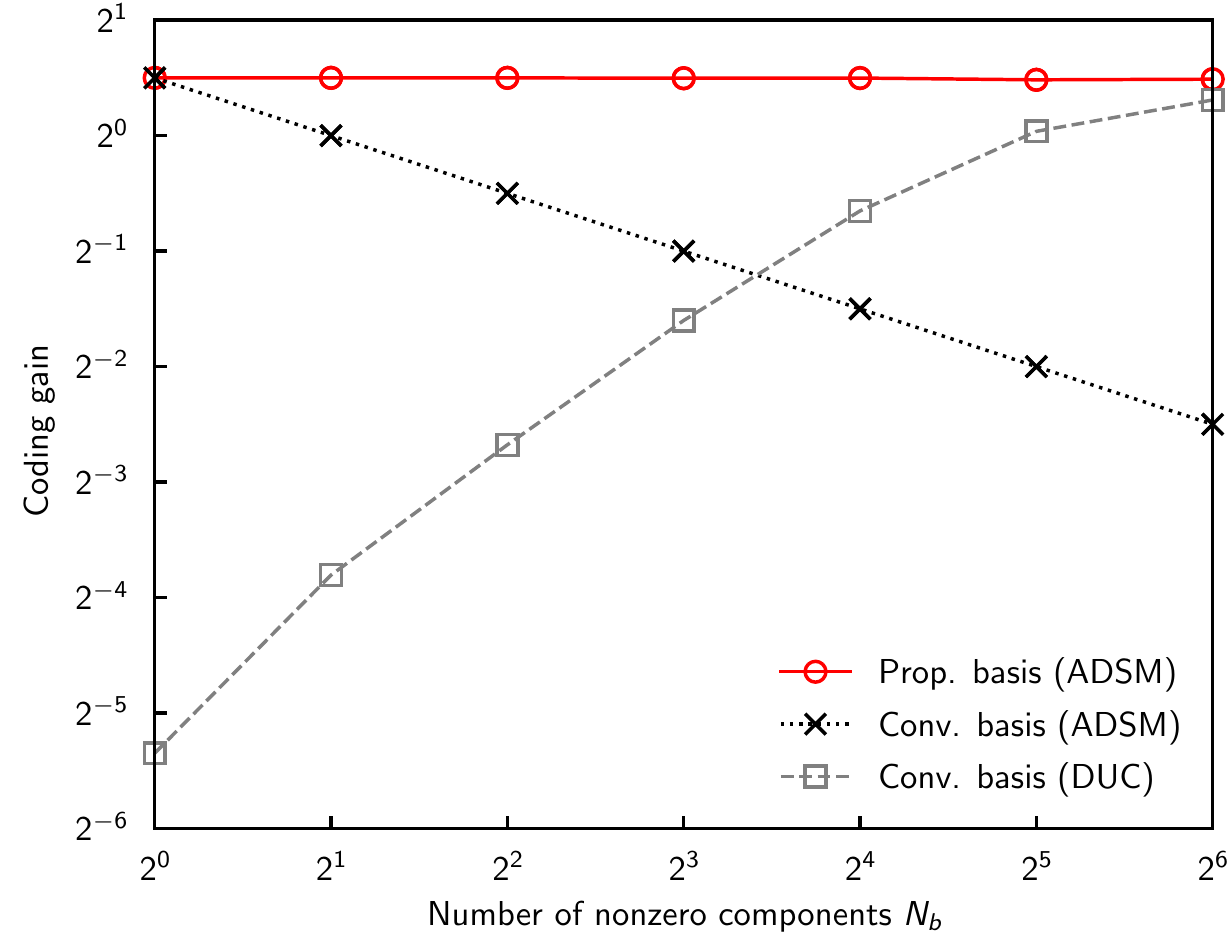}
	\caption{Comparison of coding gain for the number of transmit antennas $M=64$, the number of receive antennas $N=2$, and the length of input bit sequence $B=8$, where the number of nonzero components was varied from $N_b=2$ to $64$.}
	\label{fig:compare:coding:gain}
\end{figure}
Fig. \ref{fig:compare:coding:gain} shows the coding gain for $M=64,~N=2$, and $B= 8$, where the number of nonzero components was varied from $N_b=2$ to $64$. As shown in Fig. \ref{fig:compare:coding:gain}, the nonsquare ADSM scheme having the conventional hybrid basis \eqref{COH:hybrid:basis} exhibited the decrease in the coding gain as the number of nonzero components $N_b$ increased since the value of nonzero components of the hybrid basis must be 1. Similarly, the nonsquare DUC scheme having the conventional hybrid basis exhibited the opposite trend to the ADSM case since in DUC optimization, $u_1, u_2, \cdots ,u_M$ are optimized but only $N_b$ components are transmitted with hybrid basis. 
It was observed in Fig. \ref{fig:compare:coding:gain} that our proposed basis achieved the best coding gain for any $N_b$ as compared to the conventional basis since we optimize $N_b$ complex number for $N_b$ nonzero components. 
\begin{figure}[tb!]
	\centering
	\subfigure[AMI]{
        \includegraphics*[clip,scale=0.64]{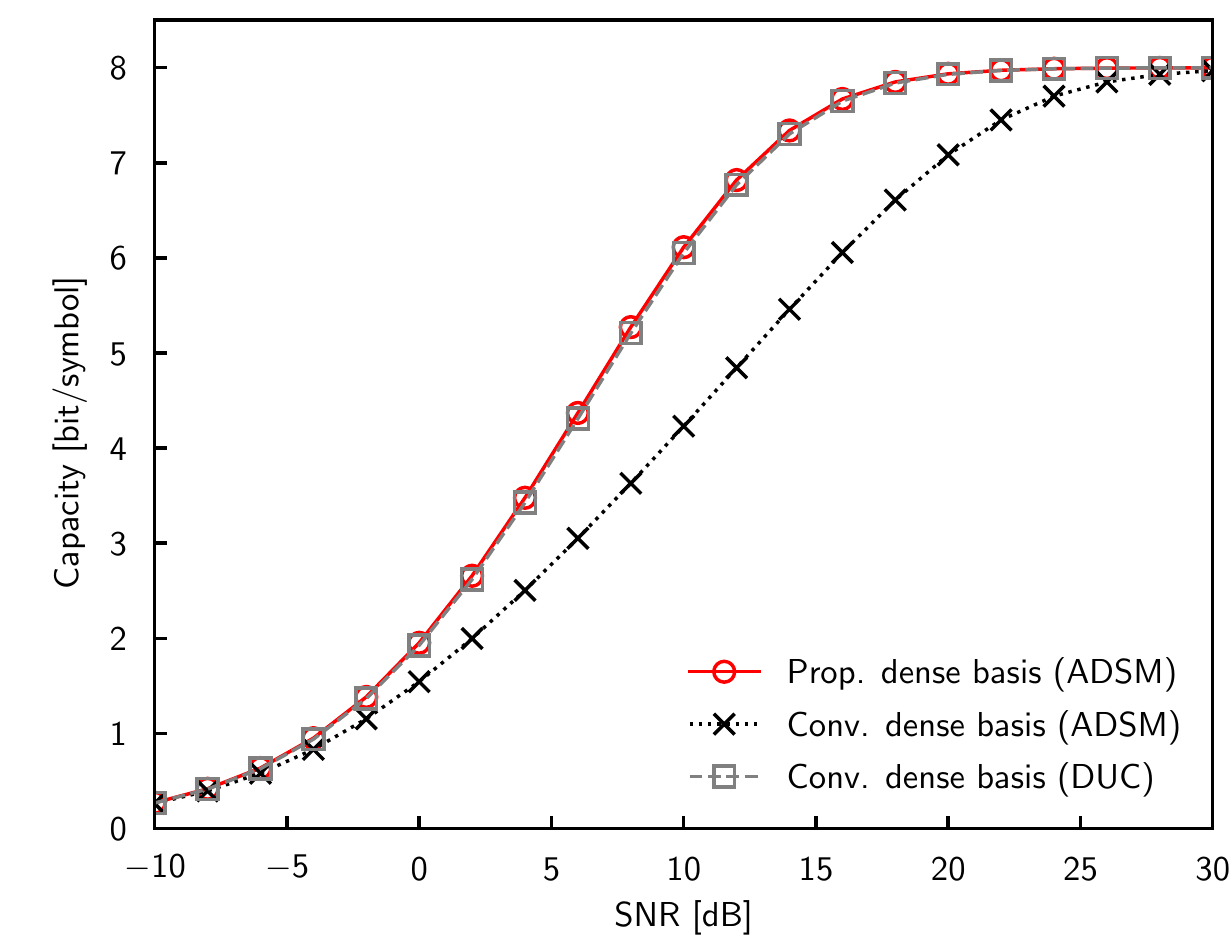}
        \label{fig:AMI}
	}
	\subfigure[BER]{
		\includegraphics*[clip,scale=0.64]{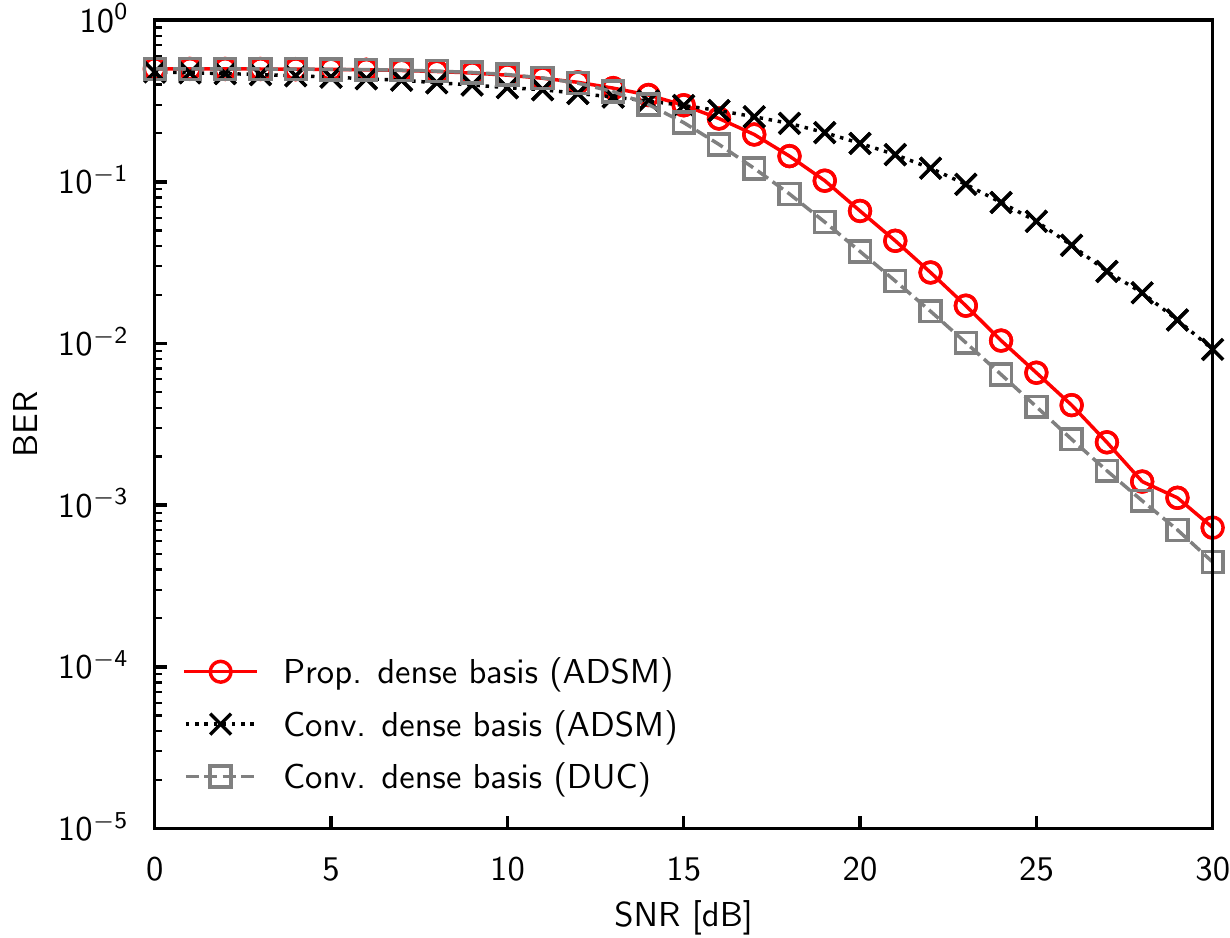}
		\label{fig:BER}
	}
	\caption{Comparisons of AMI and BER for the number of transmit antennas $M=64$, the number of nonzero components $N_b=64$, the number of receive antennas $N=2$ and the length of input bit sequence $B=8$.}
	\label{fig:compare:M64}
\end{figure}

Fig.~\ref{fig:compare:M64} shows the AMI and BER comparisons of the conventional and proposed methods, where we considered $M=N_b=64,~N=2$, and $B= 8$.
As shown in Fig. \ref{fig:compare:M64}, the proposed method succeeded in improving both AMI and BER as compared to the conventional method, and achieved the same AMI and BER as the nonsquare DUC, which required time-consuming optimization. These results were consistent with the result given in Fig. \ref{fig:compare:coding:gain}. 
It was shown in fig. \ref{fig:BER} that the BER of the proposed method was about 1 dB worse than that of the nonsquare DUC scheme, although both achieved a similar coding gain.

\section{Conclusions\label{sec:conc}}
In this paper, we proposed a low-complexity optimization method for the N-DSTC scheme, which is suitable for noncoherent massive MIMO scenarios.
The numerical comparisons demonstrated that the proposed optimization method finished 430 times faster than the conventional method, which allowed us to increase the number of transmit antennas.
In terms of the coding gain, the proposed basis outperformed the conventional hybrid basis for any numbers of transmit antennas and nonzero components, which was also verified by the AMI comparison.
In terms of BER, we observed 1 [dB] gap between the conventional nonsquare DUC and the proposed nonsquare ADSM scheme, which should be improved in future work.

\footnotesize{
\bibliographystyle{IEEEtranURLandMonthDiactivated}
\bibliography{main}
}

\end{document}